# Switchable magnetic phases in CrSBr$_{1-x}$Cl$_x$ and CrSBr/CrSCl heterostructures

Authors: Alberto M. Ruiz[1] and José J. Baldoví[1,*]

[1]Instituto de Ciencia Molecular, Universitat de València, Catedrático José Beltrán 2, 46980 Paterna, Spain. E-mail: j.jaime.baldovi@uv.es

## ABSTRACT

Strategies such as chemical substitution, strain engineering and van der Waals stacking offer powerful means to control magnetism in 2D materials, enabling the emergence of novel quantum phenomena. Here, we investigate the magnetic properties of bulk CrSBr$_{1-x}$Cl$_x$ ($x$ = 0-1) and the strain-dependent switchable magnetic phases in bilayers CrSBr, CrSCl, and CrSBr/CrSCl heterostructures, using first-principles calculations. Our findings demonstrate that Cl incorporation in CrSBr induces competing interlayer antiferromagnetic and ferromagnetic couplings, leading to a spin-glass state when the Cl content exceeds a 67%. This is accompanied by a decrease of magnetic anisotropy, where a *XY*-type magnetic ground state is observed in pristine CrSCl. For CrSBr and CrSCl bilayers, we report a reversible strain-induced switching between interlayer antiferromagnetic and ferromagnetic configurations. Furthermore, we reveal the emergence of interfacial ferromagnetism in CrSBr/CrSCl heterostructures, stabilizing an in-plane magnetization axis for CrSCl by proximity to CrSBr. These results underscore the potential of manipulating magnetism in 2D magnetic materials through precise control of magnetic phases.

## INTRODUCTION

Since the groundbreaking discovery of long-range ferromagnetic (FM) order in 2D materials such as $Cr_2Ge_2Te_6$ and $CrI_3$, research on 2D magnetism has rapidly expanded, driven by their unique properties and their potential for spintronic and magnonic applications at the limit of miniaturization.[1,2] This has led to the isolation of a diverse range of 2D magnetic materials, including the metallic ferromagnets $Fe_xGeTe_2$ (x = 3, 4, 5) and $Fe_3GaTe_2$, as well as Mott antiferromagnets like $MPS_3$ (M = Fe, Mn, Ni).[3–10] Among these, CrSBr has attracted a growing attention due to its semiconducting nature, air stability and A-type antiferromagnetic (AF) ordering with a relatively high magnetic Néel temperature ($T_N$) of 132 K.[11–13] One of the key advantages of this class of systems is their tunability by different approaches such as strain engineering, electrostatic doping, applied pressure, or molecular deposition, making them promising candidates for the precise control of magnetism at the 2D limit.[14–18]

In this context, doping has proven particularly effective to modify the properties of transition metal dichalcogenides (TMDs), even enabling a nonmagnetic to FM transition by varying the chemical compositions.[19–22] Regarding chemical substitution in 2D van der Waals magnets, the spin-orbit coupling strength has been manipulated in $CrCl_{3-x}Br_x$ upon addition of Br atoms, resulting in a transition from a *XY*- ($CrCl_3$) to an Ising-type anisotropy ($CrBr_3$).[23] Similarly, in $CrSBr_{1-x}Cl_x$ (*x* = 0-0.67), the incorporation of Cl atoms significantly influences the lattice parameters and magnetic interactions, reducing both magnetic anisotropy and Néel temperature ($T_N$), accompanied by a predicted phase transition from interlayer AF to FM ordering in CrSCl.[24] On the other hand, strain engineering is another powerful tool due to the outstanding deformation capacity of 2D materials, and it has allowed to tune their electrical, optical, superconducting and magnetic properties.[25–28] For instance, strain enhances the Curie temperature ($T_C$) and perpendicular anisotropy in $Cr_2Ge_2Te_6$,[29] while uniaxial strain exceeding 1% triggers a reversible AF-to-FM phase transition in CrSBr,[30] tuning its magnetic anisotropy, $T_C$ and magnon dynamics.[31–33] Furthermore, the weak van der Waals (vdW) interactions between adjacent layers in 2D materials facilitate the fabrication of heterostructures. This offers a platform to explore interface-driven phenomena such as interfacial ferromagnetism, as observed in $CrI_3/CrCl_3$ heterostructures,[34–37] despite both $CrI_3$ and $CrCl_3$ show an AF order in their pristine forms, or the reported magnetic hysteresis opening in twisted CrSBr monolayers.[38]

Despite these advances, a comprehensive theoretical analysis to understand the evolution of the magnetic properties in $CrSBr_{1-x}Cl_x$ (*x* = 0-1), as well as the impact of strain to control magnetism in CrSCl and CrSBr/CrSCl heterostructures is still lacking. In this work, we use first-principles calculations to investigate: (*i*) the progressive modification of the structural and magnetic properties of in $CrSBr_{1-x}Cl_x$ (*x* = 0-1), (*ii*) the strain-dependent magnetic phase transitions in CrSBr and CrSCl bilayers, and (*iii*) the emergence of interfacial ferromagnetism in CrSBr/CrSCl heterostructures. Our findings provide key insights into the design and tunability of this promising family of 2D materials for the next generation of spintronic devices.

**RESULTS AND DISCUSSION**

**Doping CrSBr$_{1-x}$Cl$_x$**

We initially performed first-principles calculations on pristine bulk CrSBr. The optimized lattice parameters ($a$ = 3.52, $b$ = 4.71 and c = 7.97 Å) align well with previous experimental and theoretical data.[13,39,40] The interlayer exchange coupling, determined by the energy difference between the FM and AF states ($\Delta E$ = $E_{FM}$-$E_{AF}$), yielded a value of $\Delta E$ = +106.84 μeV/Cr and B$_b$ = 0.6 T, where B$_b$ is the external field required to switch spins from an AF to a FM configuration along the $b$ axis, confirming that CrSBr is an A-type antiferromagnet.[41] For CrSCl, the calculated lattice parameters are $a$ = 3.42, $b$ = 4.73 and c = 7.34 Å, implying that the substitution of Br by Cl induces an anisotropic compression of $a$ and $c$ axes (arising from the smaller radius of Cl compared to Br) and a slight expansion of the $b$ lattice constant.[24] In this case, $\Delta E$ = -4.35 μeV/Cr, which results in a substantial suppression of the antiferromagnetism and an almost negligible interlayer coupling. This observation agrees with the reported magnetic properties of CrSBr$_{1-x}$Cl$_x$ ($x$ = 0−0.67), where a spin glass behaviour is observed for a Cl content in the range $x$ = 0.57−0.67, resulting from competing AF and FM interlayer exchange interactions.

To better understand the glassy behaviour, we analysed the evolution of the interlayer coupling in bulk CrSBr$_{1-x}$Cl$_x$ ($x$ = 0.00−1.00), along with changes in lattice parameters and magnetic anisotropy (Figs.1a-f). The analysis involves constructing a supercell of dimensions 2×1×2, where Br are randomly replaced by Cl atoms to evaluate the impact of increasing the Cl content on those properties (Fig. 1a). Regarding the structural features, one can observe a continuous decrease of the $a$ and $c$ lattice constants, as well as the overall volume ($V$), whereas the lattice parameter $b$ shows a slight increase upon addition of Cl (Fig. 1b). This leads to an approximate 10% reduction in $V$ for CrSCl relative to CrSBr (see Table S1 for a comparison with experimental values reported by Telford et al.[24]). In terms of the interlayer exchange interactions, we observe a progressive weakening of the AF coupling as the Cl content increases from $x$ = 0 to $x$ = 0.67 (Fig.1c), becoming almost negligible at $x$ = 0.75 ($\Delta E$ = -6.09 μeV/Cr). This picture is compatible with the abovementioned glassy behaviour that has been observed in CrSBr$_{1-x}$Cl$_x$. However, in contrast to the hypothesis of Telford et al.,[24] who suggested that the interlayer coupling of CrSBr$_{1-x}$Cl$_x$ would progressively vary upon increasing Cl content (ultimately transitioning towards a strong FM state for CrSCl), our calculations show a different trend. Specifically, beyond $x$ = 0.67, the FM coupling does not continue to strengthen. Instead, it reaches a saturation point, with the energy difference $\Delta E$ stabilizing around $\Delta E \approx 0$ μeV/Cr for compositions in the range $x$ = 0.67−1.00. This implies that CrSCl retains its spin-glass behaviour rather than transitioning to a FM state. In Fig. S1, we present additional calculations for the $\Delta E$ using an expanded supercell to include more intermediate values of $x$, which confirm the same tendency. We attribute this behaviour to two competing effects: (*i*) a preferential FM coupling between Cr that is mediated by Cl atoms, and (*ii*) a detriment of the FM interactions due to reduction in interlayer spacing as Br is progressively replaced by Cl. The substitution of Br by Cl limits the AF character exhibited by CrSBr, reaching a saturation point for $CrSBr_{0.33}Cl_{0.67}$. Beyond that, the further reduction of the interlayer spacing −driven by the smaller ionic radius of Cl− outweighs the enhancement of the FM character due to increased Cl content, ultimately resulting in an almost null interlayer coupling for CrSCl.

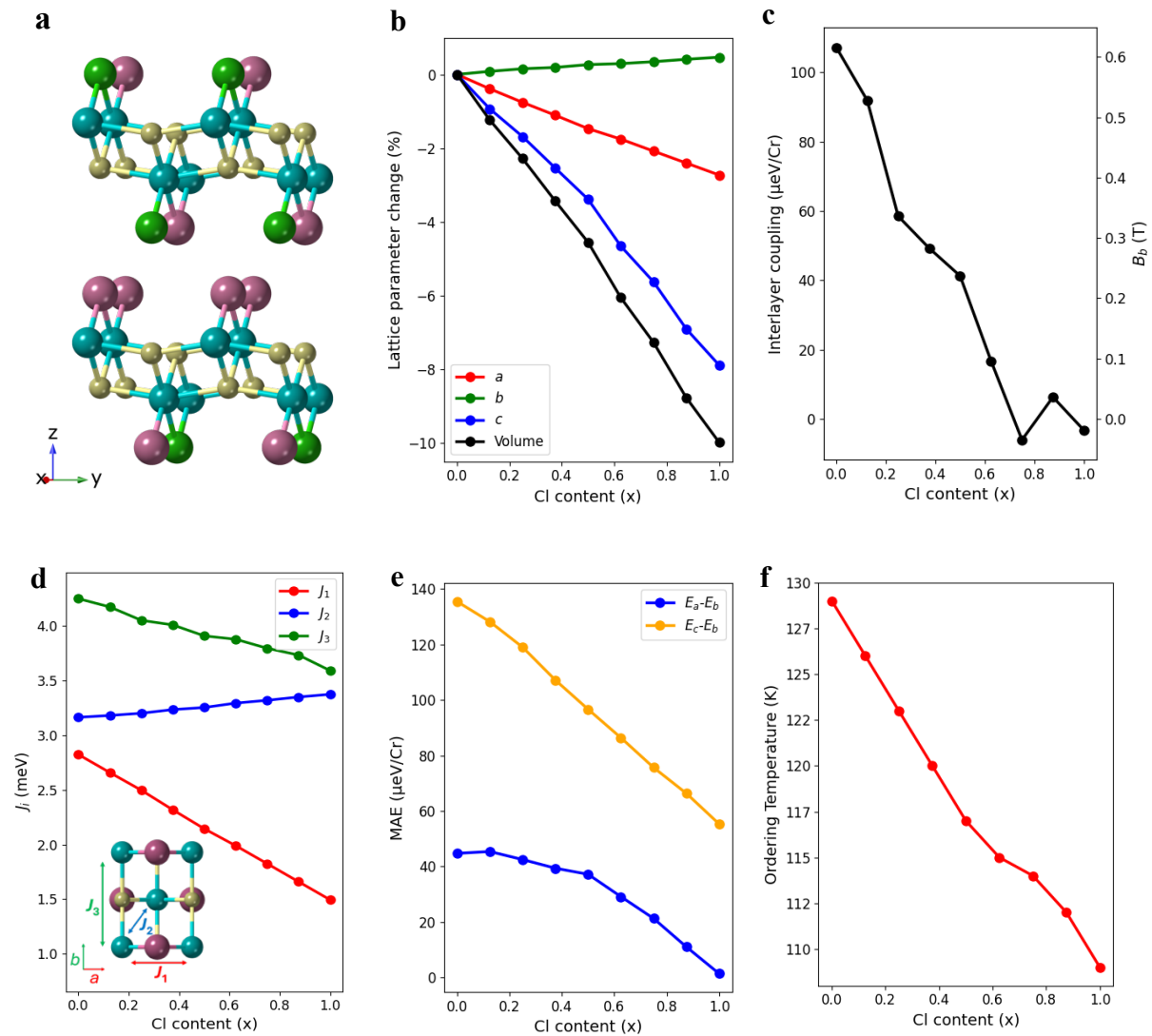

**Fig. 1. | Structural and magnetic properties of CrSBr$_{1-x}$Cl$_x$ ($x$ = 0.00−1.00). a** Lateral view of CrSBr$_{0.625}$Cl$_{0.375}$. Cyan, yellow, pink and green balls represent chromium, sulfur, bromine and chlorine atoms, respectively. **b** Evolution of lattice parameters, **c** interlayer exchange coupling (ΔE), **d** intralayer couplings J$_1$, J$_2$ and J$_3$, **e** magnetic anisotropy energy (MAE) and **f** ordering temperature for CrSBr$_{1-x}$Cl$_x$ ($x$ = 0-1).

The effect of the Cl incorporation on the intralayer magnetic exchange interactions J$_1$, J$_2$ and J$_3$ was also evaluated (Fig. 1d). These parameters correspond to the interactions between Cr atoms along *a* axis (J$_1$), diagonal direction (J$_2$) and *b* axis (J$_3$). Upon Cl substitution, we observe a continuous and significant reduction of both $J_1$ and $J_3$, while $J_2$ shows a slight increase. The reduction of $J_1$ is attributed to the compression of *a* lattice constant, which amplifies the AF contribution of $d_{xy}$-$d_{xy}$. Conversely, expansion of the *b* axis facilitates better alignment of the $d_{z2}$–$p_z$–$d_{z2}$ orbitals between Cr and S atoms, therefore enhancing the AF exchange channel of $J_3$. Finally, the enhanced FM character of $J_2$ in CrSCl is associated to an increase FM contribution of $d_{yz}$-$d_{x2-y2}$ (Fig. S2).[31]

CrSBr exhibits triaxial magnetic anisotropy, with the magnetic moments oriented along the *b* directions. The intermediate and hard magnetization states are associated with spin orientations along the *a*- and *c*-axes, respectively, showing calculated anisotropy values of 45 μeV/Cr and 135 μeV/Cr. Upon substituting Br with Cl, a decrease in the magnetic anisotropy energy (MAE) is observed across the entire composition range ($x$ = 0.00−1.00) (Fig. 1e). Unlike interlayer coupling, the reduction in MAE is progressive. Notably, the spin-orbit coupling contribution to MAE (SOC-MAE) is significantly reduced with increasing Cl content, while the shape-MAE, driven by magnetic dipole-dipole interaction, shows only minor variations (Fig. S3). This reduction is primarily attributed to the weaker SOC of Cl compared to Br. On the contrary, the magnetic moments remain almost unaltered, decreasing very slightly from 2.951 $μ_B$/Cr in CrSBr to 2.949 $μ_B$/Cr in CrSCl. Fig. 1e indicates that CrSCl shows a *XY*-type spin dimensionality, characterized by an easy magnetization plane, with the hard state corresponding to spins oriented along the *c* direction and showing an anisotropy value of 55 μeV/Cr. This behaviour is compatible with the one observed in the $CrX_3$ (X = Cl, Br, I) family, where magnetism in both $CrBr_3$ and $CrI_3$ is stabilized by an out-of-plane uniaxial anisotropy,[1,42] while in $CrCl_3$ monolayer the magnetic moments lay within the plane due to the weak SOC of Cl.[43] From the values of the exchange couplings and the MAE, we computed the ordering temperature for CrSBr$_{1-x}$Cl$_x$ (Fig. 1f). The obtained critical temperature of CrSBr is 130K, aligning perfectly with experimental findings.[11,41] By increasing the Cl content, it progressively decreases arising from the reduction of intralayer exchange couplings and MAE, resulting in a final value of 110K for CrSCl. Finally, a comparison of the electronic properties reveals that the band gap remains unchanged (Fig. S4), in line with photoluminescence measurements.[24]

**Bilayer CrSBr and CrSCl**

Then, we focus on the interaction between adjacent layers in both CrSBr and CrSCl. For such purpose, we studied the properties of the bilayer structures. Analogous to bulk, bilayer CrSBr exhibits an A-type antiferromagnet with optimized in plane lattice constants of *a* = 3.51 and *b* = 4.71 Å. The spins point along the *b* axis (Fig. 2a) with an interlayer coupling of ΔE = +69.05 μeV/Cr. This reduced interlayer coupling compared to the bulk suggests that an external field ($B_b$) of 0.4T is sufficient to switch the magnetization from AF to FM, in agreement with magneto-transport measurements.[41] Despite the decrease of the interlayer coupling, the magnetic anisotropy remains largely unchanged, with values of 36 and 160 μeV/Cr along the *a* and *c* directions, respectively. On the other hand, bilayer CrSCl possess optimized lattice parameters of *a* = 3.42 and *b* = 4.73 Å and retains a spin-glass behaviour (Fig. 2b), where ΔE = +1.07 µeV/Cr. Furthermore, the system exhibits an easy magnetization plane with the hard axis aligned along the $c$ direction, showing an anisotropy value of 84 µeV/Cr.

Due to the weak interlayer coupling of bilayer CrSCl, its magnetic ground state is highly susceptible to external perturbations. Strain engineering has been shown to effectively modulate the magnetic properties of CrSBr, where it exhibits a reversible AF-to-FM phase transition under deformations along the $a$ direction.[30] Therefore, we computed the evolution of interlayer magnetic exchange coupling in CrSBr and CrSCl as a function of strain along the $a$ and $b$ axes (Fig. 2c). Our results show that a tensile strain of 1.1% along $a$ induces an AF-to-FM switching in CrSBr, consistent with experimental strain-dependent photoluminescence measurements reporting a required strain of 1.3%.[30] Conversely, a 2% elongation of $b$ lattice constant triggers interfacial ferromagnetism. Such strain levels have been experimentally realized in other 2D materials;[44,45] therefore, we anticipate a strain-driven magnetic phase transition in CrSBr induced by deformations along of the $b$ axis. Similarly to CrSBr, tensile strain in CrSCl triggers an enhancement of FM interactions, driving the system from a glassy state towards a FM configuration with ΔE = -50 and -30 µeV/Cr at $\varepsilon_a$ and $\varepsilon_b$ = 3%, respectively. On the other hand, compressive strain favours an AF alignment for both materials. In CrSBr, the critical field needed to switch spins would increase from 0.4T to 0.9T, 1.5T and 2T at $\varepsilon_a$ = -1, -2 and -3%, respectively. Compressive strains along $b$ enhance $B_b$ to a maximum of 1.1T for $\varepsilon_b$ = -3%. For CrSCl, $B_b$ reaches a value of 0.8T (0.3T) for $\varepsilon_a$ ($\varepsilon_b$) = -3%.

The evolution of the intralayer exchange couplings under uniaxial strain reveals that $J_1$ is primarily affected by deformations along the $a$ direction, $J_3$ is mainly sensitive to distortions along the $b$ axis, while $J_2$ is influenced by both, as it takes place along the diagonal direction (see Fig. 1d and Figs. S5 and S6). Specifically, we observe that expansion (compression) of $a$ triggers an enhancement (reduction) of $J_1$, whereas the application of tensile (compressive) strain in $b$ results in weakened (strengthened) values of the ferromagnetic $J_2$ and $J_3$ (Figs. S5 and S6). On the other hand, deformations of both axes lead to changes of the MAE. In CrSBr, tensile strain along the $a$ axis significantly reduces the magnetic anisotropy, while compressive strain enhances it (Fig. S7).[32] A similar trend is observed for distortions along the $b$ direction, albeit with smaller variations. Consequently, the combination of enhanced interlayer coupling and increased anisotropy implies that the saturation fields required to switch spins from $b$ to the intermediate ($a$) and hard ($c$) directions increase upon compressive strain. On the other hand, CrSCl retains an easy magnetization plane within the studied range of strain (Fig. S8). Similar to CrSBr, deformations of $a$ axis significantly impacts the out of plane anisotropy of the system, reducing it under tensile strain. The derived critical temperatures of bilayer CrSBr and CrSCl are 136 K and 112 K, respectively, which are enhanced up to 145 K and 136 K upon elongation of $a$. Similarly, compressions in $b$ result in enhanced critical temperatures up to a maximum of 146 K and 131 K (Fig. S9).

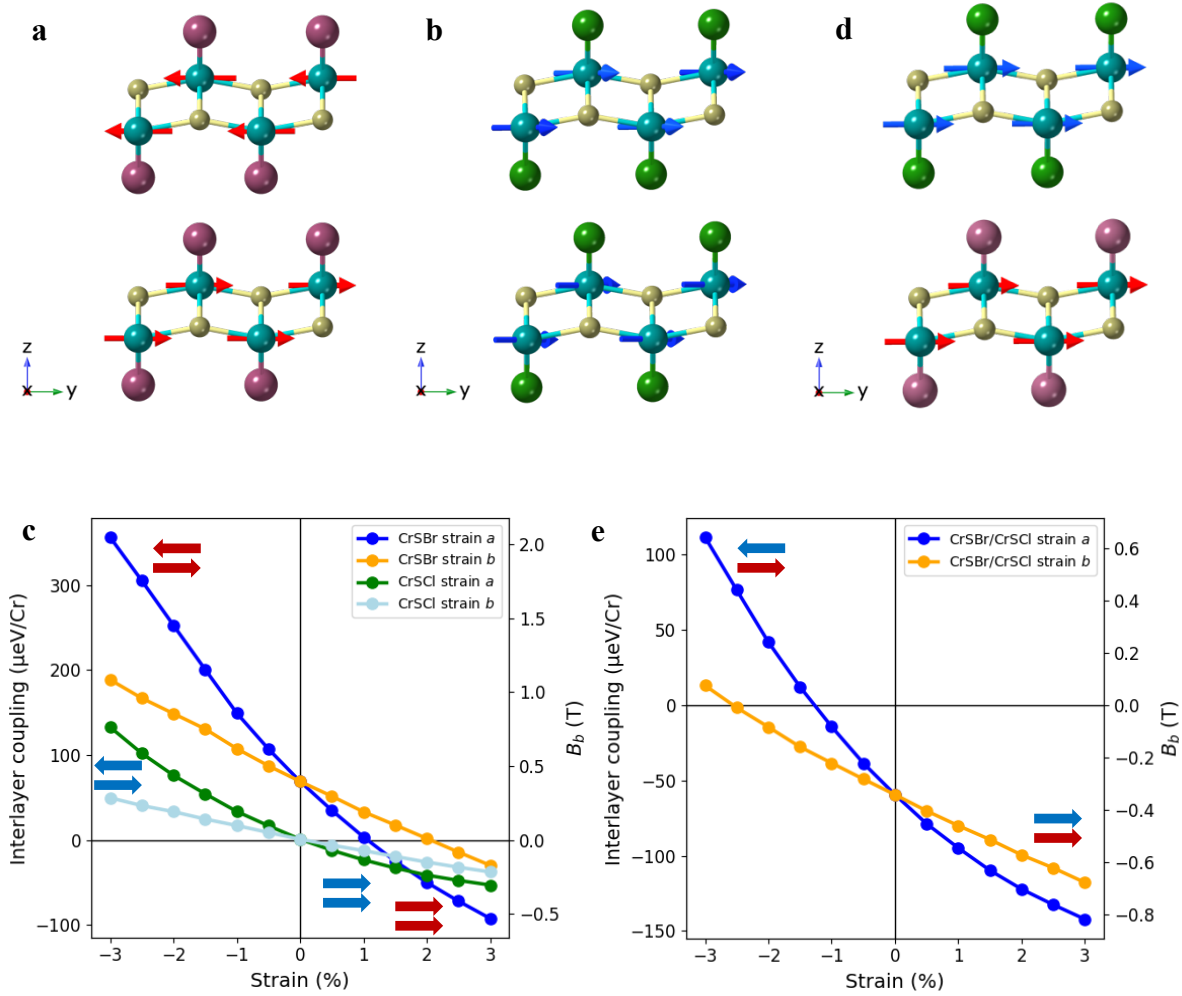

Fig. 2. | **Magnetic properties of CrSBr, CrSCl bilayers and a CrSBr/CrSCl heterostructure**. **a, b** Lateral view of bilayers CrSBr and CrSCl. Red (green) arrows represent the favourable orientation of spins for CrSBr (CrSCl). In CrSCl each layer has an easy magnetization plane along with a negligible interlayer coupling. **c** Evolution of the interlayer exchange coupling for bilayers CrSBr and CrSCl upon applied strain. **d** Lateral view and **e** strain-dependent interlayer exchange coupling of a CrSBr/CrSCl heterostructure.

## 1L CrSBr/ 1L CrSCl heterostructure

We next investigated the interfacial properties of a 1 L CrSBr/ 1L CrSCl heterostructure. Our results reveal the presence of interfacial ferromagnetism between both materials (ΔE = -59.41 μeV/Cr), with their magnetic moments oriented along the *b* axis (→→), resulting in a non-zero net magnetization (Fig. 3d). Note that in →→, the first and second spin correspond to the CrSBr and CrSCl layers, respectively. Other magnetic configurations were also tested, including (*i*) in plane AF coupling (→←), (*ii*) perpendicular antiferromagnetism (↑↓) and (*iii*) a configuration where the magnetic moments of one layer point in plane while the other are oriented along the *c* axis (→↑), among others (Table S2). However, these configurations were found to be less stable. A similar behaviour has been reported for a $CrI_3$/$CrCl_3$ heterostructure, where interfacial ferromagnetism is observed despite both materials exhibit interlayer AF coupling in their pristine bilayer forms.[37] The calculated anisotropy of the

CrSBr/CrSCl is 27 and 108 µeV/Cr for the *a* and *c* axes, respectively. These values indicate a reduction in the anisotropy in CrSBr due to its proximity to CrSCl. Additionally, the latter now experiences an effective anisotropy field influenced by both the interfacial ferromagnetic interaction and the intrinsic anisotropy of CrSBr. As a result, the previously discussed *XY*-type spin state observed in pristine CrSCl is suppressed, turning to a triaxial anisotropy behaviour with spins pointing along the *b* axis.

Regarding the evolution of the interlayer coupling upon strain, we observe that a compression of *a* induces a phase transition from →→ to →← at $\varepsilon_a$ = -1.2%. Further compression enhances the AF character, resulting in a strong AF state with $B_b$ = 0.6T at $\varepsilon_a$ = -3%. On the other hand, compressive strain along *b* has a more subtle impact on the interlayer coupling, where the parallel and antiparallel configurations are nearly degenerate at $\varepsilon_b$ = -3%. In the CrSBr/CrSCl heterostructure, the critical temperature is 136 K, being almost identical to the one of CrSBr bilayer. This observation is corroborated by calculations of the exchange coupling within the heterostructure, which reveal that the intralayer couplings for each CrSBr layer remain almost unchanged despite its proximity to the CrSCl layer (Table S3). Furthermore, the strain-dependent variation of the intralayer exchange couplings, MAE and $T_C$ follows the same tendency as in the pristine bilayer structures (Figs. S10-13).

**2L CrSBr/ 1LCrSCl heterostructure**

We then examined the magnetic properties of a heterostructure formed by a CrSBr bilayer in proximity to a CrSCl monolayer in order to assess for the impact of the CrSCl/CrSBr interface on the interlayer coupling between adjacent CrSBr layers. In the pristine state, trilayer CrSBr displays AF interlayer coupling, with ΔE = +86.01 µeV/Cr and spins pointing along the *b* axis ($B_b$ = 0.5 T). Similar to the bilayer case, trilayer CrSBr undergoes an AF-to-FM magnetic phase transition under tensile strains of $\varepsilon_a$ = +1 % and $\varepsilon_b$ = +2 % (see Fig. S14). In the 2L CrSBr/1L CrSCl heterostructure, the magnetic moments align along the *b* direction, with the CrSBr bilayer exhibiting AF coupling. The CrSCl monolayer, in turn, is ferromagnetically aligned with the adjacent CrSBr layer, resulting in a ←→→ configuration (Fig. 3a). Here, the first two spins correspond to CrSBr bilayer, and the third to the neighbouring CrSCl monolayer. This observation suggests that the interfacial antiferromagnetism between CrSBr layers remains unaffected by their proximity to CrSCl. We also explored alternative magnetic configurations, including (*i*) ferromagnetic coupling between the CrSBr with an antiferromagnetic alignment to CrSCl (→→←), (*ii*) antiferromagnetic coupling across all the three layer (←→←) and (*iii*) a global ferromagnetic configuration (→→→). However, those were found to be less stable (Table S4).

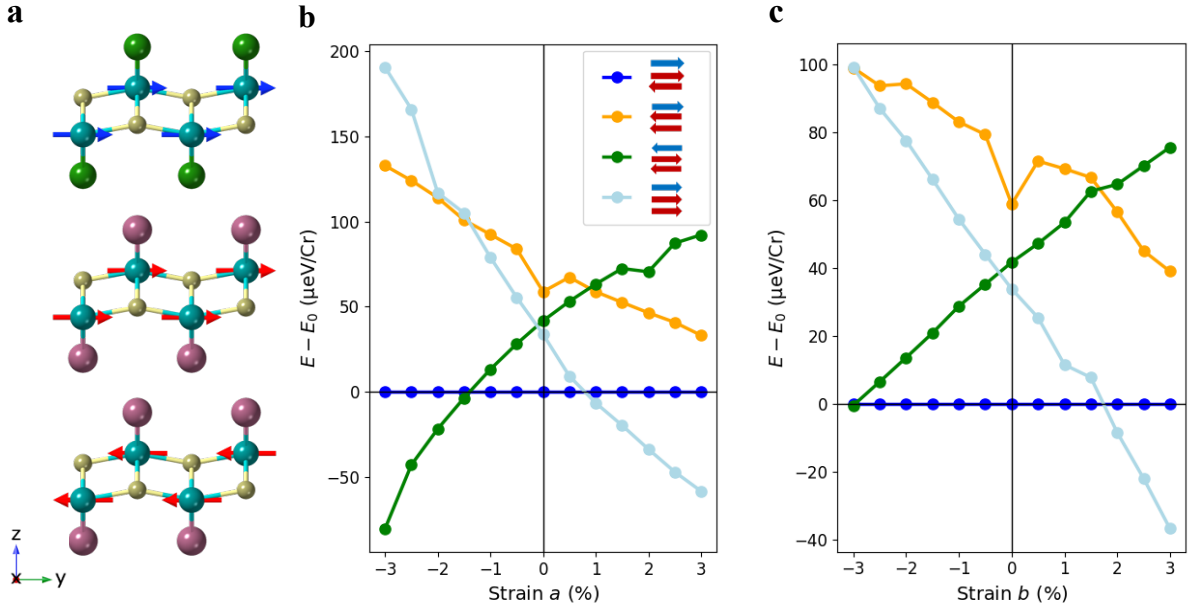

**Fig. 3 | Magnetic properties of a 2L CrSBr/1L CrSCl heterostructure. a** Lateral view of 2L CrSBr/ 1L CrSCl heterostructure. **b, c** Relative stability of the different magnetic phases compared to the ground state as a function uniaxial strain along *a* and *b*, respectively.

Figs. 3b and 3c depict the relative stability of the different magnetic phases compared to the ground state (←→→) as a function of ε. Upon elongation, we observe a gradual stabilization of the fully FM configuration (→→→), leading to a phase transition from ←→→ to →→→ at $\varepsilon_a = 0.75$ % and $\varepsilon_b = 1.75$ %. This behaviour mirrors the strain response observed in pristine bilayer and trilayer CrSBr, where tensile strain induces an AF-to-FM phase transition between adjacent CrSBr layers. However, for the pristine systems, the phase transition occurs at higher strain values of $\varepsilon_a = +1$ % and $\varepsilon_b = +2$ % (see Figs. 2c and S11). Therefore, while contacting CrSCl with CrSBr layers does not modify the interfacial antiferromagnetism of the later, it is weakened compared to the pristine form. On the other hand, a compression of the lattice destabilizes the FM alignment between CrSCl and CrSBr layers, eventually transitioning towards an AF coupling (←→←) at $\varepsilon_a = -1.50$ % and $\varepsilon_b = -3$ %.

**Conclusions**

In summary, we investigated the effects of chemical substitution, strain engineering, and heterostructures design on the magnetic properties of CrSBr using first-principles calculations. Our work provides a detailed understanding of the progressive evolution of magnetic properties in CrSBr$_{1-x}$Cl$_x$ ($x$ = 0-1), revealing the emergence of a spin-glass behavior, *XY*-type anisotropy and reduced Néel temperature in CrSCl. We demonstrated that the interlayer magnetic exchange coupling in CrSBr and CrSCl can be selectively tuned along the crystallographic directions (*a* and *b*) as a function of strain, enabling reversible switching between ferromagnetic and antiferromagnetic configurations. Moreover, we reported the emergence of interfacial ferromagnetism in CrSBr/CrSCl heterostructures and proved the effectiveness of strain engineering in tailoring their magnetic ground state. These findings underscore the versatility of modulating magnetism and its significance for the implementation of 2D magnetic materials with precise control of magnetic phases in versatile spintronic devices.

**METHODS**

We carried out spin polarized density-functional theory (DFT) calculations using the VASP package[46]. The generalized gradient approximation (GGA) was employed as the exchange-correlation functional. To account for dispersion interactions, we employed the nonlocal functional optB86b, and the plane-wave cutoff was set to 500 eV. The calculations for bulk CrSBr and CrSCl were performed considering cells of dimensions 1×1×2 to account for the antiferromagnetism between adjacent layers. The Brillouin zone was sampled by a fine Γ-centered 20×20×2 k-point Monkhorst–Pack. The simulations of the doped $CrSBr_{1-x}Cl_x$ systems were performed creating supercells of dimensions 2×1×2, therefore a total of 8 Br atoms can be substituted progressively by Cl. In that case, we adopted a Monkhorst–Pack $k$-mesh of 10×20×2. To properly describe the bilayer and trilayer systems, a 18 Å vacuum space perpendicular to the layers was employed and the Brillouin zone was sampled by a fine Γ-centered 20×20×1 k-point Monkhorst–Pack. A tight binding model was constructed based on maximally localized Wannier function as implemented in the Wannier90 code[47] with a reduced basis set formed by the $d$ orbitals of Cr atoms and the $p$ orbitals of Br, Cl and S. The magnetic couplings were obtained using the Green's function method using the TB2J code[48] using supercells of dimensions 20×20×1. For the calculations of the intralayer exchange couplings we considered an effective Hubbard $U_{eff}$ = 3 eV. The critical temperature was obtained by performing atomistic simulations implemented in the VAMPIRE package[49] employing systems of dimensions 15 nm × 15 nm, with both equilibration and averaging phases done using 5,000 steps.

## ACKNOWLEDGEMENTS

The authors acknowledge the financial support from the European Union (ERC-2021-StG101042680 2D-SMARTiES). J.J.B acknowledges the Generalitat Valenciana (grant CIDEGENT/2023/1) and A.M.R. thanks the Spanish MIU (Grant No FPU21/04195). The calculations were performed on the HAWK cluster of the 2D Smart Materials Lab hosted by the Servei d'Informática of the University of València.

## COMPETING INTERESTS

The authors declare no competing interests.